\documentclass[pra,aps,superscriptaddress,twocolumn,floatfix,notitlepage,longbibliography]{revtex4-2}

\usepackage{amsmath}
\usepackage{amsfonts}
\usepackage{amssymb}
\usepackage{bm}
\usepackage{mathrsfs}
\usepackage{graphicx}
\usepackage[svgnames]{xcolor}
\usepackage{xspace}
\usepackage{stmaryrd}
\usepackage{laussy}
\usepackage[sort&compress]{natbib}

\colorlet{RED}{red}

\begin{document} 
\title{Correlations in Circular Quantum Cascades}

\author{Miguel \'Angel Palomo Marcos}
\affiliation{Departamento de Física Teórica de la Materia Condensada, Universidad Autónoma de Madrid, 28049 Madrid, Spain}

\author{Eduardo Zubizarreta Casalengua}
\affiliation{Walter Schottky Institute, School of Computation, Information and Technology and MCQST, Technische Universit\"at M\"unchen, 85748 Garching, Germany}

\author{Elena del Valle}
\affiliation{Departamento de Física Teórica de la Materia Condensada, Universidad Autónoma de Madrid, 28049 Madrid, Spain}
\affiliation{Condensed Matter Physics Center (IFIMAC), Universidad Autónoma de Moadrid, 28049 Madrid, Spain}
\affiliation{Institute for Advanced Study, Technische Universität München, 85748 Garching, Germany}

\author{Fabrice P.~Laussy}
\email{fabrice.laussy@gmail.com}
\affiliation{Instituto de Ciencia de Materiales de Madrid ICMM-CSIC, 28049 Madrid, Spain}

\date{\today}

\begin{abstract}
  We introduce a one-way, one-quantum cascade, whereby a single
  excitation proceeds one-directionwise in a ladder of energy levels.
  This makes a variation from more famous two-way cascades where the
  excitation can go up and down following its excitation or relaxation
  in the ladder. We provide closed-form solutions for two-photon
  correlation functions between any transitions in such circular
  cascades. We discuss how the rich correlations that result from what
  appears to be an extremely simple implementation, are essentially
  those which have been obtained from complex
  architectures relying on strongly-correlated, many-body physics or
  cavity QED effects, and might constitute the liquid phase of the new
  family of time materials.
\end{abstract}

\maketitle

\section{Introduction}

\begin{figure*}
  \includegraphics[width=.8\linewidth]{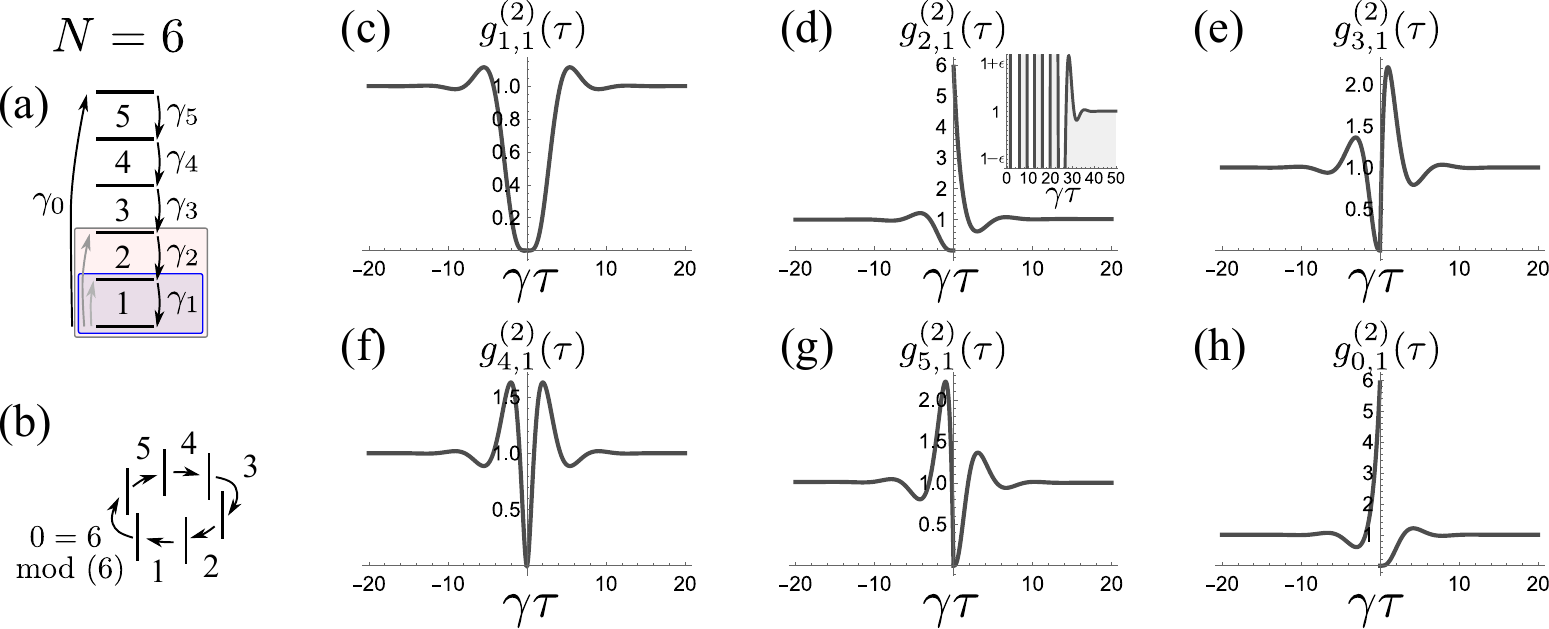}
  \caption{(a) $N$-level cascade for~$N=6$, with five radiative
    transitions at rates~$\gamma_i$ ($1\le i\le 5$) plus one
    reloading~$\gamma_0$. The two-level system is in the dark-shaded
    box ($N=2$) and the two-photon cascade from a three-level system
    in the light-shaded box ($N=3$), with their own~$\gamma_0$
    transition.  The generalization to any~$N$ is obvious.  (b) The
    equivalent circular representation, where pumping becomes a
    regular transition and could also be monitored. (c--h) All the
    possible correlation functions for~$N=6$, when
    all~$\gamma_i=\gamma$. There are~four different traces, modulo
    time mirror symmetry, that account for all the 36 possible
    correlations. The inset of $g^{(2)}_{2,1}$ in~(d) makes a zoom in
    the window~$1\pm\epsilon$ where~$\epsilon=10^{-6}$, allowing to
    resolve five of the oscillations from the cascade, which carry on
    forever and are present in all correlation functions
    for~$N\ge3$. \label{fig:Sat13Jul122521CEST2024}}
\end{figure*}

Cascading is a widespread phenomenon that can amplify features of a
system, as is best illustrated by the domino effect.  With bosons, it
led to the idea of the quantum cascade laser~\cite{kazarinov71a},
whose realization~\cite{faist94a} opened new perspectives and regimes
of operation for coherent light~\cite{gao23a}. The variation where the
active medium itself is also bosonic~\cite{liew13b} led to new regimes
of superbunched emission~\cite{liew16a}. Here, we consider cascades of
a single quantum of excitation, down a ladder of~$N$ levels. In
contrast to the previous cascades, both the active medium and the
radiation field thus remain at the level of single quanta. There is a
rich variety of platforms where this can take
place~\cite{crubellier87a,briegel96b,rodnyi00a,bunzli07a,zhou12a,peng15a,fritzsche21a}
and such cascading is therefore not new, being in fact basically
intrinsic to the way optical emitters release their
excitation~\cite{fontana_book82a}. Phenomena like quantum cutting,
whereby one quantum of excitation 
results in the production of several photons, i.e., with quantum
efficiency larger than~1, have been long
known~\cite{sommerdijk74a,piper74a}.  The main variation here will
come from bringing such cascades in a stationary regime.  When a
cascade can be maintained in a steady state, it may result in a chain
reaction that gives rise to new dynamical regimes, as in the
aforementioned quantum-cascade and bosonic lasing.  Under Continuous
Wave (CW) excitation, cascades have been particularly studied by the
semiconductor community to characterize spectral lines from complex
multi-excitonic states~\cite{bayer00a} through the study of their
correlations. This allows to identify the order of
transitions~\cite{dekel00a,regelman01a} and measure their radiative
lifetimes~\cite{camhyval70a}. The technique has been for instance
demonstrated for the characterization of the tri-exciton, with photon
cascades involving up to~$N=5$ excitonic levels, with three radiative
steps~\cite{persson04a} or four photon transitions from
quadexcitons~\cite{arashida11a}. In the latter case and in other
striking examples (e.g., with the tri-exciton again, but correlating
all transitions simultaneously~\cite{schmidgall14a}), this was under
pulsed excitation so that correlations were reduced to coincidences
(in particular, of the bunching type), and thus deprived from the time
dynamics, which had to be complemented by time-resolved
photoluminescence. The reason for pulse excitation is that such
cascades balance two types of transitions: downward as the system
releases its excitation, and upward as it gets re-excited by the
constant driving. As a result, depending on the pumping power, one
gets stuck at more or less high stages of the ladder. Consequently,
correlations are more easily obtained between consecutive steps, where
the system is pinned, although photon correlations being so robust to
low signal, they have also been successfully demonstrated between
far-apart transitions even in the CW
regime~\cite{molas16a}. Nevertheless, this two-way option for the
excitation which can hop up and down at any stage, weakens the cascade
as a whole and effectively turns it into a succession of two-photon
cascades.  Here, we draw attention to spectacular features present in
one-way cascades where the excitation can only relax downward, until
it reaches the ground state, at which point it gets excited again to
the top, as sketched in Fig.~\ref{fig:Sat13Jul122521CEST2024}(a).
Recently, there has been a surge of interest for such unidirectional
flow of various types of
excitations~\cite{corzo19a,liedl24a,pasharavesh24a}. Notably, both
quantum-cascade lasers and bosonic cascades maximize their properties
when the flow is unidirectional. Here, we focus on the simplest
possible one-way cascading: incoherent pumping initiates the cascade
at random times by resetting the single excitation to the highest
level~$N$ from the bottomest one.  There needs not be an actual top
and bottom levels, and the structure could really be circular, as
sketched in Fig.~\ref{fig:Sat13Jul122521CEST2024}(b). Given that this
better captures the structure of the transitions, we shall refer to
such cascades as ``circular''. We will briefly discuss ways to achieve
them but we first motivate their interest, for the correlations
between the photons they produce.

\section{Liquid light and time crystals}

Our motivation is a previous report that
such (circular) photon cascades with a large number of intermediate
steps endow autocorrelations in time that are reminiscent of those
found spatially in liquids~\cite{zubizarretacasalengua24a}, and that
familiar single-photon correlations from two-level systems are a
particular case of this broader cascading scheme.  This shows that
cascading is interesting even if one step only of the cascade is
radiative.  Interestingly, in the limit of a large number of levels,
this approaches perfect single-photon sources with opening of a time
gap~\cite{khalid24a}. Although stationary, such sources exhibit
features of pulsed emission, but without any external synchronization.
Because this occurs without any coherence and/or periodic driving,
this can be seen as self-oscillations~\cite{jenkins13a}, although of
correlations from a system that is itself stationary.  We became
latterly aware that such peculiar correlations had been previously
predicted for strongly-interacting Rydberg atoms maintained in their
electromagnetically-induced transparency
configuration~\cite{otterbach13a}. Such correlations are not available
to Kerr-type, point-interacting photon blockade. Instead, they follow
from a quite dramatic phase-transition from the optically dense active
medium itself, interacting strongly and nonlocally~\cite{sevincli11a}
and clustering into small self-avoiding regions, related to Wigner
crystallization~\cite{otterbach13a}. Switching off the driving field
transfers the spatial correlations to temporal photons.  This
reinforce one's feeling that such correlations imprinted into the
optical field correspond to a new phase (here we cannot write ``of
matter'' since that is now more general than that).  A remarkable
point is that such correlations---which have been described as
``nonlinear quantum optics in its extreme, in which individual photons
behave as impenetrable particles''~\cite{chang08a}---are
straightforwardly sculpted by a mere cascade mechanism, itself excited
incoherently and thus with no external coherence, order or
synchronicity. This produces an infinitely-long stream of
quasi-crystalline order (to take the terminology of the Rydberg
effect) as opposed to a finite size, quenched pulse in most
configurations (Zeuthen \emph{et al.}~\cite{zeuthen17a} discuss the CW
driving converting the Poisson input into a regular pulse train of
single photons).  Photon liquefaction (to take the cascade
terminology) thus appears to be more general and fundamental than one
could think since it occurs in completely unrelated platforms from
completely different mechanisms.

Another and maybe even closer concept from both the phenomenological
and lexical points of view, is that of a time
crystal~\cite{sacha_book20a}.  It posits that breaking the
time-translation symmetry should result in the spontaneous emergence
of a repeating period for the ground state, something ``perilously
close to perpetual motion'', as it was qualified in both the
back-to-back seminal papers~\cite{wilczek12a,shapere12a}. This was
initially proposed for closed and conservative systems, which
triggered a series of no-go theorems~\cite{bruno13c,watanabe15a}. The
idea for a counterpart in time of something so fundamental and
commonplace in space was so compelling that despite various proofs of
its impossibility, it triggered multiple quests to overcome its
mathematical restrictions. A first way out was found by turning to
coherently-driven, out-of-equilibrium systems, in which case, a weaker
breaking of temporal invariance becomes possible, giving rise to
Discrete Time Crystals~\cite{sacha15a,else16a,khemani16a}, where the
new (spontaneous) period is now a multiple of that of the
driving. Whereas Wilczek's time crystal remains elusive to this day,
even theoretically, discrete time crystals have been readily
demonstrated experimentally~\cite{zhang17a,choi17b}. Then, it was
furthermore realized that a much-closer version to the original
proposal was also possible in open, non-conservative
systems~\cite{iemini18a,buca19a,kessler19a}, giving rise to Continuous
Time Crystals.  When the dissipation comes from a continuous, but
incoherent, pumping, this introduces no frequency in the system and
any periodic response breaks time-translation symmetry.  This stronger
scenario was also quickly realized
experimentally~\cite{kongkhambut22a,carrarohaddad24a}. Our circular
cascades exhibit several traits that one could attribute to time
crystals, and our terminology of time
liquefaction~\cite{zubizarretacasalengua24a} already arrived to
similar conclusions, although distinctions are as many as
resemblances.  In particular, our system exhibits oscillations without
periodic driving. The spontaneous occurence of its time ordering is
thus defined only by the microscopic details of the system. But,
instead of periodic, infinitely-lived oscillations, that are
characteristic of a crystal, circular cascades exhibit slowly drifting
and decaying ones, indeed better corresponding to a liquid. Our system
is also temporal only with no spatial dimension and, in particular,
does not involve neither many-body interactions nor the thermodynamic
limit.  This may still constitute a new addition to the family of
``time materials''~\cite{verstraten21a} in the form of what Wilczek
himself described as hypothetical ``time liquids'' along with also
time quasicrystals and time glasses~\cite{wilczek19a}.

Regardless of how liquid light (defined as light whose Glauber's
two-photon correlation is the same, in time, as the pair-correlation
function, in space, of liquids) relates, or not, to time matter (in
the sense of Wilczek), it seems clear that it should be present in
various configurations or in disguise, as it represents a fundamental
phase of the optical field, which should be further scrutinized,
especially as it better corresponds to what one understands as a
single-photon source~\cite{khalid24a}. For instance, this type of
correlations is also possible in coherently-driven few-step
cascades~\cite{ooi07a}. The cascade implementation might thus be a
privileged platform to realize and investigate it, given its
simplicity in both conceptual, theoretical and applied aspects, at
least when contrasted to the Rydberg blockade version, which is a
highly sophisticated platform, replete with complications in
real-laboratory implementations (such as ``pollutants'', i.e., Rydberg
atoms unduly excited by their peers in the propagating
blockade~\cite{bienias20a}).

In the following, we generalize our previous treatments of such
circular cascades by studying all the two-photon correlations (not
only one transition in isolation~\cite{zubizarretacasalengua24a}) as
well as correlations from any subset of transitions. These are other
degrees of freedom from our scheme not available from the Rydberg
gas. Since large cascades are naturally more difficult to realize than
small ones, and because two-photon cascade from a three-level system
has already enjoyed considerable popularity and attention, we also
focus on this particular case.

\section{Formalism}

Our system of study is simple. If one contents with correlations of
the transitions alone, then all the results can be obtained through
rate equations, as was also the case for the other types of
cascades~\cite{pegg86a,dekel00a}.  Such equations are best written in
their matrix form for the probabilities~$p_i$ of each level~$\ket{i}$
to be excited at time~$t$ for~$1\le i\le N$:
\begin{equation}
  \label{eq:Thu25Jul181335CEST2024}
  \partial_t\begin{pmatrix}
    p_{N}\\p_{N-1}\\\vdots\\p_2\\p_1
  \end{pmatrix}
  =
  \begin{pmatrix}
    -\gamma_{N-1} & \hdots & & & \gamma_0 \\
    \gamma_{N-1} & -\gamma_{N-2} & & & \\
    \vdots & \ddots & \ddots & \\
     & & \gamma_2 & -\gamma_{1} \\
    0 & & & \gamma_1 & -\gamma_0
  \end{pmatrix}
  \vec p\,,
\end{equation}
where~$\vec p$ is the vector on the lhs.  This makes for $j$
transitions, from level~$\ket{j+1}$ to level~$\ket{j}$
for~$1\le j\le N-1$, occuring at rate~$\gamma_j$, while the reloading
brings level~$\ket{1}$ back to level~$\ket{N}$ at
rate~$\gamma_0$. Consequently, there are as many transitions as
levels, although the reloading one is usually not regarded as a
radiative transition, but as an excitation. This is an indication of
the conceptual difference between the model (of levels) and the
observable (transitions).  In a quantum treatment, $p_i$ would be
obtained as $\bra{i}\rho\ket{i}$ from the density matrix~$\rho$ of the
system, but in absence of genuinely quantum observables, we do not
need a master equation. Such a quantum description is eventually
desirable, but even there, it would be especially because one should
correlate the emitted photons instead of the transition
operators~$\ketbra{j}{j+1}$ from which such photon supposedly
originate. We write ``supposedly'' because there is some inherent
quantum uncertainty as to the identity of a \emph{detected} photon,
whose attributes (such as frequencies, time of detection, etc.) might
not be sufficient to distinguish it from other measured events.  By
correlating transition operators, one makes the assumption that
photons are clearly distinguishable, which can be the case, at least
in some approximation, for instance when different transitions differ
greatly in frequencies. Note that frequencies do not even appear
explicitly at our current level of description. A physical
photodetection description which does not make such approximations
requires a theory of frequency-resolved photon
correlations~\cite{delvalle12a}, and we postpone this to a future
work. Here, at the level of transition operators, we can solve
Eq.~(\ref{eq:Thu25Jul181335CEST2024}) with standard linear algebra,
expressing the general solution as ``basically a sum of exponentials
with different time constants''~\cite{persson04a}
\begin{equation}
  \label{eq:Tue30Jul130337CEST2024}
  \vec p(t)=\sum_{j=0}^{N-1}C_j\vec A_je^{\lambda_jt}
\end{equation}
where~$\vec A_j$ and~$\lambda_j$ are the eigenvectors and eigenvalues of the
relaxation matrix in Eq.~(\ref{eq:Thu25Jul181335CEST2024}). The
eigenvalue problem leads to the following equation:
\begin{equation}
  \label{eq:Tue25Jun113756CEST2024}
  \prod_{i=0}^{N-1}(\lambda+\gamma_i)=(-1)^{N-1}\prod_{i=0}^{N-1}\gamma_i\,.
\end{equation}
For low~$N$, we can assume all these variables to be independent, but
for large~$N$, since we do not know a closed-form solution for this
equation, to simplify our discussion and stick to the main points, we
will assume all rates (including the pumping rate), to be the same,
equal to~$\gamma$.  This is with some loss of generality, and there
are probably important qualitative effects to be found in the more
general configurations.  In this~$\gamma_i=\gamma$ approximation for
all~$i$, the eigenvalues can be found as $\lambda_j=\gamma(1-z_N^j)$
for~$0\le j\le N-1$ while the elements of the eigenvectors are given
by $A_{jk}=z_N^{jk}$ where
\begin{equation}
  \label{eq:Tue25Jun114435CEST2024}
  z_N\equiv\exp\left({2i\pi\over N}\right)
\end{equation}
is the $N$th root of unity. The steady state is proportional to the
eigenvector with eigenvalue~$0$, which always exists since the
determinant of the matrix is zero. Thus, with~$\lambda_0=0$
and~$A_{0k}=1$, also imposing~$\sum_{i=0}^{N-1}{p_i}=1$,
then~$p_k^{\mathrm{SS}}=1/N$ for all~$k$. This is the expected result
on physical grounds under the given approximations. To compute
correlation functions---the quantities of interest in this text---we
rely on the stationarity of the signal, making~$\vec p$ independent
of~$t$ and thus providing the correlations as
\begin{equation}
  \label{eq:Tue30Jul161920CEST2024}
  g_{m,n}^{(2)}(\tau)={p_{n|m-1}(\tau)\over p_n^\mathrm{ss}}
\end{equation}
where~$p_{n|m-1}(t)=p_n(t)$ as given by
Eq.~(\ref{eq:Tue30Jul130337CEST2024}) with the initial
condition~$p_i(0)=\delta_{i,m-1}$, i.e., expressing the probability of
finding the system in state~$n$ at time~$t+\tau$ given that it started
in state~$m-1$ at time~$t$, i.e., it just underwent the transition~$m$
at this time. To compute the constant~$C_j$, we thus have to solve the
corresponding Eq.~(\ref{eq:Tue30Jul130337CEST2024}) at~$t=0$, i.e.,
\begin{equation}
  \label{eq:Tue30Jul132537CEST2024}
      \sum_{j=0}^{N-1}C_jA_{jn}={\delta_{n,m-1}}\,.
\end{equation}
The solution can be found by noting that the $N$th roots of unity sum
up to zero, or, using the formula for the sum of the first $N$ terms
of a geometric series, we find~$C_j=z_N^{-j(m-1)}/N$. This provides
us with the general correlation function for~$\tau\ge0$ between any
two transitions:
\begin{equation}
  \label{eq:Tue25Jun114035CEST2024}
  g^{(2)}_{m,n}(\tau)=1+\sum_{j=1}^{N-1}z_N^{j(n-m+1)}\exp\left(-\gamma\tau(1-z_N^j)\right)\,.
\end{equation}
We restricted to positive~$\tau$, while for negative~$\tau$:
\begin{equation}
  \label{eq:Tue9Jul141459CEST2024}
   g^{(2)}_{m,n}(\tau)=g^{(2)}_{n,m}(-\tau)\,.
\end{equation}
This is a generalization of 
our previous autocorrelation function~\cite{zubizarretacasalengua24a}
when~$m=n$ (which was obtained with the waiting time distribution), in
which case the level indices disappear and all transitions feature the
same correlations, that are also symmetric in time. The case~$m\neq n$
describes cross-correlations, between contiguous, or not,
transitions. This describes the density of probability that if a
photon is detected from transition~$m$, then another will be (or was
if~$\tau<0$) detected from transition~$n$ at time~$\tau$.  The
case~$N=1$ washes out all correlations and correspond to classical
(Poisson) emission (of the ideal gas in the perspective of
Ref.~\cite{zubizarretacasalengua24a}). The case~$N=2$ is the
incoherently-pumped two-level system, that describes the paradigmatic
CW single-photon source. Although a cascading scheme, it is not
traditionally regarded as such but as a single-transition quantum
emitter, although its statistics, in particular its failure to
suppress efficiently multiphoton emission, is best understood from the
circular cascade~\cite{khalid24a}, as we will further discuss.  The
case~$N=3$ corresponds to the two-photon cascade from a three-level
system, which also received a fair share of attention, and to which we
return in more details in Section~\ref{sec:Sat29Jun131254CEST2024}.
The case~$N=6$ is shown in Fig.~\ref{fig:Sat13Jul122521CEST2024} as a
particular case of the obvious general~$N$. We cover this case first
to provide results of general validity.

\section{Multilevel cascade}
\label{sec:Sat29Jun201742CEST2024}

The $N^2$ possibilities~Eq.~(\ref{eq:Tue25Jun114035CEST2024}) for
correlating any two transitions from a $N$-level cascade, assuming
equal rates, reduce to $N$ different positive~$\tau$ traces for the
correlation functions which we define as:
\begin{equation}
  \label{eq:Tue9Jul153153CEST2024}
    g^{(2)}_{k\{N\}}(\tau)\equiv 1+\sum_{j=1}^{N-1}z_N^{jk}\exp\left(-\gamma\tau(1-z_N^j)\right)\,.
\end{equation}
The relationship between all the possible
transitions~(\ref{eq:Tue25Jun114035CEST2024}) and
this backbone
structure~(\ref{eq:Tue9Jul153153CEST2024}) is obtained by rotation of
the cascade as, for any~$1\le k,l\le N$:
\begin{equation}
  \label{eq:Sat13Jul125215CEST2024}
  g^{(2)}_{k,l}=g^{(2)}_{(l-k+1)\{N\}}
\end{equation}
where on the left-hand side, the two indices now correspond to which
transitions are correlated, while on the right-hand side, one has
Eq.~(\ref{eq:Tue9Jul153153CEST2024}) where the curly bracket index
serves to track the number of transitions. These functions further
combine to provide the full~$\tau$ (positive or negative) correlations
according to Eq.~(\ref{eq:Tue9Jul141459CEST2024}), so there are
finally~$\lfloor N/2\rfloor+1$ different all-$\tau$ traces for
a~$N$-level circular cascade, taking into account time mirror symmetry
(the~$+1$ to account for the autocorrelation).  The properties of the
correlations thus reduce to those of
Eq.~(\ref{eq:Tue9Jul153153CEST2024}). By cyclicity of the roots of
unity, $g^{(2)}_{k\{N\}}=g^{(2)}_{(k+N)\{N\}}$.  All display maximum
antibunching
\begin{equation}
  \label{eq:Sat13Jul153352CEST2024}
  g^{(2)}_{k\{N\}}(0)=0
\end{equation}
except if~$k=0$ (or a multiple of~$N$) in which case, $z_N^0=z_N^N=1$
and the bunching is equal to the number of levels:
\begin{equation}
  \label{eq:Tue9Jul155646CEST2024}
  g^{(2)}_{0\{N\}}(0)=N\,.
\end{equation}
For convenience, we use the terminology of ``bunching'' and
``antibunching'' for both auto and cross-correlations, although this
properly describes autocorrelations only. 

\begin{figure}
  \includegraphics[width=.8\linewidth]{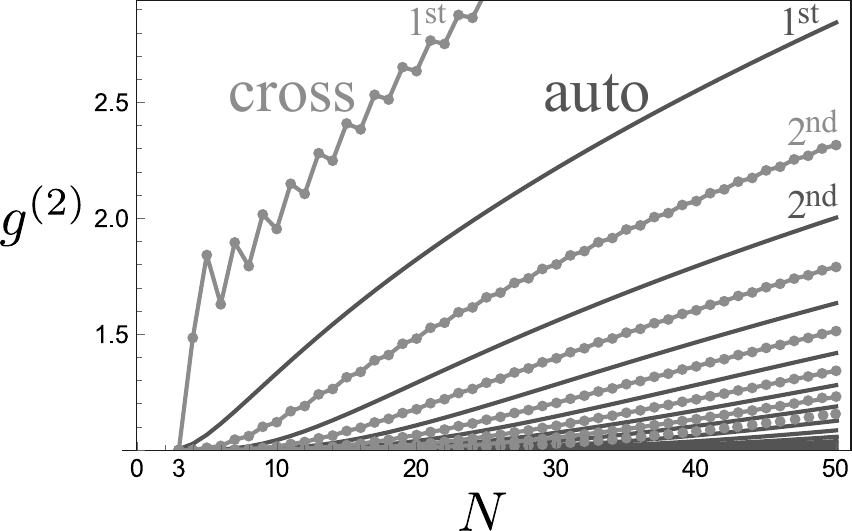}
  \caption{Magnitudes of the successive peaks of~$g^{(2)}_{1\{N\}}$
    (autocorrelations, dark lines)
    and~$g^{(2)}_{1\{\lfloor{(N+3)/2}\rfloor\}}$ (cross-correlations
    between opposite transitions in the ladder, dotted lines) as a
    function of~$N$. The cross-correlations oscillate because of their
    asymmetry for odd~$N$ (the largest peak is taken). For clarity, only
    the first seven peaks of cross-correlations are retained.
    \label{fig:Sat3Aug113209CEST2024}}
\end{figure}

The above description means that it is enough to know the
correlation between one transition (say the bottom one) and the
successive ones until halfway trough the ladder to know all the
correlations (involving a time mirror symmetry for the remaining
transitions). For instance, in Fig.~\ref{fig:Sat13Jul122521CEST2024},
all autocorrelations are identical to~$g^{(2)}_{1,1}$ in panel~(c)
and~$g^{(2)}_{2,1}$ is the time-mirror of~$g^{(2)}_{0,1}$, or,
alternatively, is identical to~$g^{(2)}_{1,0}$. The physical
interpretation of such symmetries is clear: $g^{(2)}_{2,1}$ shows the
density of probability to detect a photon from the transition~1 after
detecting a photon from transition~2. Given that these transitions are
contiguous, the chances are high as directly related to the transition
rate. The probability is further boosted by the knowledge of the
current state of the system, which could otherwise be in any of
the~$N$ states prior to effectuating a transition, but removing this
uncertainty makes it~$N$ time more likely that it will now emit at the
sought transition, thus explaining the integer factor for the
autocorrelation~(\ref{eq:Tue9Jul155646CEST2024}).  There is also a
notable revival of probabilities, in the form of an elbow
at~$\tau\approx N/\gamma$. This corresponds to the second-order
detection where the system went around through the whole cascade again
after emitting the first photon, before detecting the second one.
There are similar peaks at~$\tau\approx kN/\gamma$ for all~$k$,
corresponding to going round the whole cascade $k$~times before
detection. One can resolve them by zooming in the correlation
function, as shown in the inset of~$g^{(2)}_{2,1}$ in
Fig.~\ref{fig:Sat13Jul122521CEST2024}(d), or by plotting~$g^{(2)}-1$
in log-scale. Such oscillations occur in all correlation functions
for~$N>2$ and have been described in more details for the
autocorrelation function~\cite{zubizarretacasalengua24a}. Their
magnitude increases slowly but surely with~$N$, as shown in
Fig.~\ref{fig:Sat3Aug113209CEST2024} through the values of the
successive peaks of~$g_{i,i}^{(2)}(\tau)$ (dark solid lines) and
of~$g_{i,\bar\imath}^{(2)}(\tau)$ (grey dotted lines) as a function
of~$N$ for~$i$ any transition and~$\bar\imath$ the one most distant
from it in the cascade.  That is to say, the successive maxima
of~$g_{1\{N\}}^{(2)}$ and~$g^{(2)}_{1\{\lfloor{(N+3)/2}\rfloor\}}$ are
read from a vertical cut for a given~$N$.

Much structure can be extracted by decomposing the correlation
functions as a train of spontaneous emissions, whereby the~$m$th
peak~$g_m(\tau)$ follows from the waiting time distribution~$w$ of the
corresponding event.  For instance, writing the autocorrelation
function as
\begin{equation}
  \label{eq:Thu5Dec145746CET2024}
  g_{1\{N\}}^{(2)}(\tau)=\sum_{m=1}^\infty g_m(\tau)
\end{equation}
we have in the Laplace space~$g_m(s)={N\over\gamma}w_N(s)^m$
with~\cite{zubizarretacasalengua24a}
\begin{equation}
  \label{eq:Thu5Dec143230CET2024}
  w_N(s)
  =\left(\gamma\over\gamma+s\right)^N
\end{equation}
so that, by inverse Laplace transform, we obtain the expression for
the~$m$th peak (which is neatly resolved in this form if peaks do not
overlap appreciably):
\begin{equation}
  \label{eq:Thu5Dec150457CET2024}
  g_m(\tau)=N{(\gamma\tau)^{mN-1}\over(mN-1)!}\exp(-\gamma t)\,.
\end{equation}
Remarkably, this is simply $N$-times the Poisson distribution (with
parameter~$\gamma\tau$) for the integer~$mN-1$. With hindsight, it
makes sense that the~$m$th photon from a circular cascade with~$N$
levels should involve this distribution. We remind, however,
that~$g_m$ is not a probability distribution but a component of a
correlation function. The heralding picture, however, offers such a
probabilistic connection, similarly to
Eq.~(\ref{eq:Tue30Jul161920CEST2024}). This allows us to derive
important characteristics of the peaks, such as their area, which is
the same for all peaks and equal
to~$\int_0^\infty g_m(\tau)\,d\tau=N/\gamma$. The peak maximum is
found for~$\tau_m$ such that~$g'_m(\tau_m)=0$, which is~
\begin{equation}
  \label{eq:Sat7Dec165837CET2024}
  \tau_m={mN-1\over\gamma}
\end{equation}
giving the aucorrelation~$g^{(2)}_{1\{N\}}$ at the $m$th peak
as~$g_m(\tau_m)$ or, in good approximation using Stirlings' formula,
\begin{equation}
  \label{eq:Thu5Dec151538CET2024}
  g_m(\tau_m)
  \approx{N\over\sqrt{2\pi(mN-1)}}\,.
\end{equation}
This is accurate for the well-separated peaks or early oscillations
where~$g_m(\tau_m)>1$, whereas this approximation tends to zero for
large~$m$.  In this range of validity,
Eq.~(\ref{eq:Thu5Dec151538CET2024}) recovers the solid lines in
Fig.~\ref{fig:Sat3Aug113209CEST2024}, which have been obtained from
the exact expression~(\ref{eq:Tue9Jul153153CEST2024}).
Substituting~$m$ in Eq.~(\ref{eq:Thu5Dec151538CET2024}) for~$\tau_m$
also gives the upper envelope of the autocorrelation function
as~$N/\sqrt{2\pi\tau}$.
The time~$\tau_m$ at which the $m$th peak reaches its maximum can also
be found---for~$mN$ large enough---from the mean time~$\bar\tau_m$
of~${\gamma\over N}g_m(\tau)$ turned into a density of probability for
the emission of the~$m$th photon of the cascade:
\begin{equation}
  \label{eq:Fri6Dec194743CET2024}
  \bar\tau_m\equiv{\gamma\over N}\int_0^\infty\tau g_m(\tau)\,d\tau={mN\over\gamma}\,.
\end{equation}
This gives us the standard deviation for the $m$th peak as
\begin{equation}
  \label{eq:Fri6Dec202406CET2024}
  \sigma_m^2\equiv{\gamma\over N}\int_0^\infty(\tau-\bar\tau_m)^2g_m(\tau)\,d\tau={m N\over\gamma^2}\,.
\end{equation}
Since Eq.~(\ref{eq:Thu5Dec150457CET2024}) is not numerically
stable for large~$mN$, all of the above allows us to approximate
it to:
\begin{equation}
  \label{eq:Sat7Dec163757CET2024}
  g_m(\tau)\approx{N\over\sqrt{2\pi(mN-1)}}\exp\left(-{(\tau-\bar\tau_m)^2\over2\sigma_m^2}\right)\,.
\end{equation}
For~$N$ large enough, the ratio of the separation between successive
peaks to their standard deviation is therefore of the order of
$(\bar\tau_m-\bar\tau_{m-1})/\sigma_m=\sqrt{N\over m}$.
%
%
This shows that the ``crystallization'' is very slow and the train of
pulses melts rapidly into oscillations that themselves quickly
dissolve into stationarity. The condition for at least~$k$ standard
deviations in total (${k\over2}\sigma_m$ for both peak, each with its
standard deviation~$\sigma_m$ and~$\sigma_{m+1}$) to fit in the
interval between two successive peaks is fulfilled for the~$m$ first
peaks when
\begin{equation}
  \label{eq:Sat7Dec144456CET2024}
  m\le\left\lfloor{(k^2-4N)^2\over (4k)^2N}\right\rfloor\,.
\end{equation}
This is independent of~$\gamma$.  For $k=7$ ($3.5$~standard deviation
on each side of each peak, including~$>99.97\%$ of their respective
area), this requires a~$N=72$ cascade for the first peak only to
satisfy this condition and~$N=122$ for the first two peaks.
Although neatly separated pulses require large~$N$, the oscillations
are easily formed.  From the exact results, we find that one 
requires~$N=6$ to get a bunching of the first oscillation
of~$g^{(2)}\approx 1.1$ ($10\%$ deviation,
cf.~Fig.~\ref{fig:Sat13Jul122521CEST2024}(c)) and~$N=13$ for the
second peak to reach that value. At~$N=50$, the seventh peak
has~$g^{(2)}\approx 1.13$ and the eighth one~$g^{(2)}\approx 1.09$ so
at least eight-order cascades can be considered neatly resolvable,
i.e., from the steady stream, strong correlations are maintained
between any given photon and its eighth descendant round the cascade,
each generation undergoing fifty transitions. More precise experiments
could track down such correlations arbitrarily down the stream for
any~$N$.
A similar dynamics occurs for negative times: the chances get strongly
suppressed that a photon from the transition~$1$ be observed before
($\tau<0$) one from transition~$2$, which constitutes a transition in
the wrong time order. In a physical model, such a process becomes
possible due to detector uncertainties. Here, it is strongly
suppressed because it requires to go round the full cascade to get
another photon to pause a wrong-order one. This is impeded by all the
intermediate steps. The succession of antibunching at negative~$\tau$
followed by bunching at positive~$\tau$ defines the archetype of a
photon cascade, although this is one particular step only of the
general dynamics, but an important one to which we come back in
Section~\ref{sec:Sat29Jun131254CEST2024}. The identity
between~$g^{(2)}_{2,1}$ and~$g^{(2)}_{1,0}$ is also clear as we
consider the same processes in different positions of the ladder, as
is obvious on the circular
picture~Fig.\ref{fig:Sat13Jul122521CEST2024}(b) of the cascade, in
contrast to the linear one~(a).  Similar dynamics explain
non-contiguous transitions, e.g., $g^{(2)}_{3,1}$ in panel~(e) shows
the bunching at positive~$\tau$ of the almost contiguous cascade, with
only one intermediate transition to make, whose interruption also
explains the $\tau=0$ antibunching in this and all other
non-contiguous correlations. Importantly, this flattens in time as the
distance between transitions increases in the ladder. This is the
principle of photon liquefaction, which endows the correlations with
an intrinsic, spontaneous mechanism for local time-ordering.  The
small~$\tau$ behaviors can be characterized
through a series expansion of Eq.~(\ref{eq:Tue9Jul153153CEST2024}),
but since this removes the modularity of~$k$, one must
ensure that~$1\le k\le N$:
\begin{equation}
  \label{eq:Wed10Jul120100CEST2024}
    g^{(2)}_{k\{N\}}(\tau)={N\over(N-k)!}(\gamma\tau)^{N-k}+o(\tau^{N-k+1})\,.
\end{equation}
%
The case~$k=N$ (which identifies to~$k=0$ modulo~$N$ in
Eq.~(\ref{eq:Sat13Jul125215CEST2024}) and corresponds to contiguous
transitions) is expanded not around zero (as per
Eq.~(\ref{eq:Sat13Jul153352CEST2024})) but around~$N$ (as per
Eq.~\ref{eq:Tue9Jul155646CEST2024}) and one must thus treat this case
separately, to find:
\begin{equation}
  \label{eq:Wed10Jul120905CEST2024}
  g^{(2)}_{N\{N\}}(\tau)=N\exp(-\gamma\tau)+o(\tau^{N+1})
\end{equation}
so that to next-leading order in small times,
$g^{(2)}_{N\{N\}}(\tau)=N(1-\tau)+o(\tau^2)$, in agreement with
Eq.~(\ref{eq:Wed10Jul120100CEST2024}). All together, this shows that
high-order corrections account for the long-term behaviour of these
functions, following the plateau, or exponential decay for~$k=N$, when
oscillations kick in before converging
to~$g^{(2)}_{k,N}(\infty)=1$. The small-time approximations also
confirm that the farther the transitions in the cascade, the more
suppressed is the possibility of a coincidence in the way accounted
for by cumulative random events~\cite{khalid24a}: the probability to
detect photons from two given transitions $k$-steps away is given by
the compound probabilities of the intermediate steps in the
sequence. This also ``crystallizes'' the chance of the coincidence to
occur at the $k$th trial, so one has a steeper and larger bunching for
small~$k$. Once such a one-cascade loop has been completed, the
process can then repeat, diluting the features as they become both
less strong and less well-resolved in time due to accumulations of
time uncertainties, but still resulting in all-time oscillations,
already for the~$N=3$ cascade (but not for~$N=2$, showing that the
effect is not trivial and some type of minimum cooperativeness is
required for it to occur).
\begin{figure}
  \includegraphics[width=\linewidth]{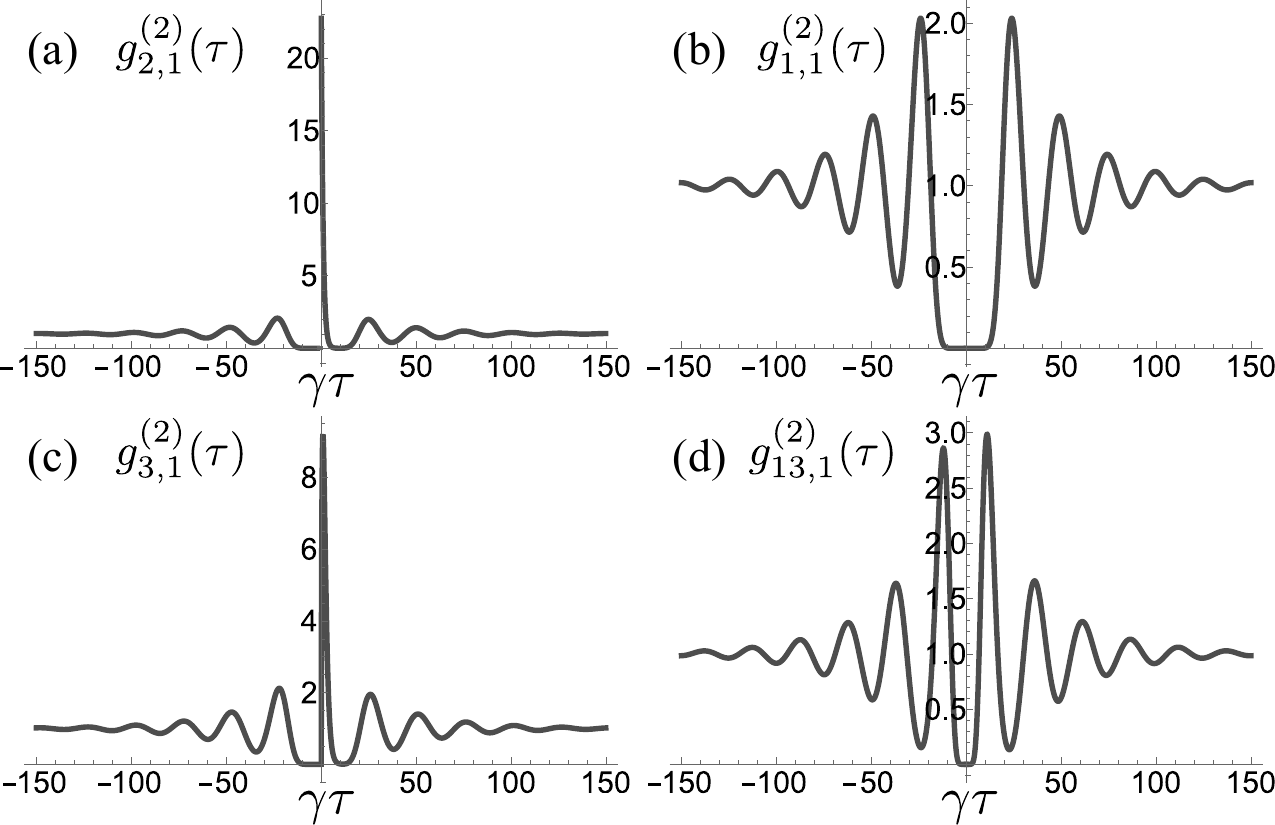}
  \caption{$N$-level cascade with~$N=25$, showing four illustrative of
    the thirteen different correlations out of the 625 possible ones,
    namely, (a) contiguous or (c) next-to-contiguous transitions as
    well as (b) autocorrelations and (d) mid-ladder
    transitions. Long~$\tau$ oscillations are now neatly resolved,
    manifesting a strong local time-ordering. The correlation function
    is discontinuous in~(a) but not in the other panels, where it
    exhibits a more or less flat plateau around~$\tau=0$ as specified
    by
    Eq.~(\ref{eq:Wed10Jul120100CEST2024}). \label{fig:Sat13Jul152249CEST2024}}
\end{figure}
An illustrative case is shown in Fig.~\ref{fig:Sat13Jul152249CEST2024}
for the case~$N=25$. In this case, there
are~$\lfloor 25/2\rfloor+1=13$ different traces but it is enough to
consider a few illustrative cases, namely, the contiguous or almost
contiguous transitions, shown in Panels~(a) and~(c), which one can
contrast with their lower~$N$ case counterparts in
Fig.~\ref{fig:Sat13Jul122521CEST2024}(d) and~(e), to see how the
liquefaction phenomenology superimposes itself to the immediate
cascading. The traces in~(a) and~(c) are very similar, except at
small~$\tau$, where the strong bunching is curbed by the impossibility
of a concidence from the in-betweener,
cf.~Eq.~(\ref{eq:Sat13Jul153352CEST2024}). As a result, the curve in
panel~(a) is discontinuous at~$\tau=0$ while that in~(c) is not. They
differ more from the autocorrelation function~$g^{(2)}_{1\{25\}}$ in
panel~(b) as well as the mid-ladder case~$g^{(2)}_{13\{25\}}$ in~(d)
that correlates the most distant transitions in the ladder. Since~$N$
is, in this case, odd, there is one more transition on one side of the
cascade and the function is not exactly symmetric
(cf.~Fig.~\ref{fig:Sat13Jul122521CEST2024}(f), which is
symmetric). Such opposite-transition correlations look qualitatively
like autocorrelations (especially when exactly symmetric) although
they are more correlated, with steeper and higher bunching peaks,
except that their temporal gap is shorter, being~$N/(2\gamma)$ or half
the gap of the~$N/\gamma$ autocorrelation, since it takes that time
only to connect the two events.  This shows how auto-correlations bear
more constrains as applying within the same stream as opposed to
cross-correlations that involve two signals. This relates to an
important concept, which has accompanied photon cascades since their
early days, namely, that of their violation of classical
inequalities~\cite{loudon80a}, predominantly, Cauchy-Schwarz
inequalities of the
type~$g^{(2)}_{nn}(0)g^{(2)}_{mm}(0)\ge[g^{(2)}_{nm}(\tau)]^2$, which
classical signals must satisfy. At our current level of description,
the violation is trivially enforced for small times, in the form (from
Eq.~(\ref{eq:Wed10Jul120100CEST2024}))
\begin{equation}
  \label{eq:Tue30Jul185833CEST2024}
  0\ge {N^2(\gamma\tau)^{2(N-m+n-1)}\over[N-m+n-1]!^2},
\end{equation}
which can never be satisfied (i.e., Cauchy--Schwarz inequalities are
always violated). The violation is larger, the farther away the
transitions.  Such violations, which indicate that one is dealing with
a quantum signal with no classical equivalent, were in fact first
observed by Burnham and Weinberg~\cite{burnham70a} with parametric
down conversion and provided the earliest experiment evidence of
non-classicality of light, well ahead of the more famous antibunching
experiment of Kimble~\emph{et al.}~\cite{kimble77a}, which is the
degenerate $n=m$ version. Circular cascades thus provide a platform to
evidence even more forcefully nonclassical correlations from light,
even though from a more mundane relaxation, with no photon splitting.

\begin{figure}
  \includegraphics[width=.9\linewidth]{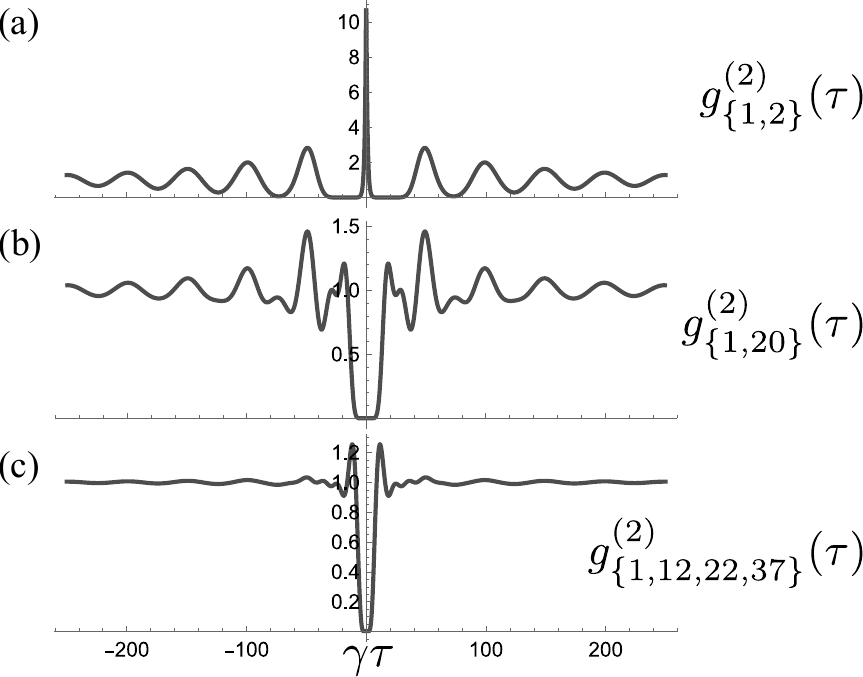}
  \caption{Many-transition correlations where all the labelled
    transitions in the subscript set are jointly detected, here for
    an~$N=50$-level cascade. (a) Correlating two consecutive
    transitions realizes a two-photon bundle liquid, with the
    appearance of a superbunching peak
    (with~$g^{(2)}_{\{1,2\}}(0)=12.5$ from
    Eq.~(\ref{eq:Fri2Aug205123CEST2024})). (b) Correlating distant
    transitions bring chaotic looking patterns. (c) A random subset of
    four transitions wash out the oscillations while still maintening
    the temporal gap, realizing a dense temporal gas. Many other
    combinations can be realized and remain to be
    investigated. \label{fig:Sat3Aug130801CEST2024}}
\end{figure}

Now that we characterized all possible transitions, we can take the
next engineering step afforded by such cascades, by collecting light
from a given subset~$\mathcal{S}=\{i_1,\cdots,i_n\}$ of
transitions. This could be enforced by having such transitions have
the same frequency and filtering out the undesired ones, whose
participation to the cascade is still necessary but not their
contribution to the emitted light. One could also think of selection
by polarization, non-radiative process or any other trick whose end
result is to collect only photons from the subset. In this case, the
correlation function follows from the basic correlation functions
(\ref{eq:Tue25Jun114035CEST2024}) as
\begin{equation}
  \label{eq:Fri2Aug201942CEST2024}
  g^{(2)}_{\mathcal{S}}(\tau)={1\over(n_\mathcal{S})^2}\sum_{i,j\in\mathcal{S}} g^{(2)}_{i,j}(\tau)
\end{equation}
where~$n{_\mathcal{S}}\equiv\#\mathcal{S}$ is the cardinality
of~$\mathcal{S}$, i.e., the number of transitions contributing to the
emission. Thanks to Eq.~(\ref{eq:Tue9Jul141459CEST2024}),
$g^{(2)}_{\mathcal{S}}$ is $\tau$-symmetric, as should be for a
correlation function. Some examples are shown in
Fig.~\ref{fig:Sat3Aug130801CEST2024}, this time for~$N=50$. Of
particular interest is to select $k$ successive transitions, in which
case, as long as~$k\ll N$, one has essentially the same
autocorrelation but pierced through by a central superbunching peak,
of magnitude
\begin{equation}
  \label{eq:Fri2Aug205123CEST2024}
  g^{(2)}_{\llbracket 1, k \rrbracket}(0)={N(n{_\mathcal{S}}-1)\over n{_\mathcal{S}}^2}\,,
\end{equation}
as shown in Fig.~\ref{fig:Sat3Aug130801CEST2024}(a). This corresponds
to a liquid of $k$-photon bundles, where each emission consists of a
so-called bundle (group) of~$k$ photons, i.e., closely-packed photons
in time as compared to the photons from the other bundles. Thanks to
the cascaded emission process, those multiphoton bundles do not have
the harmonic progression of their cavity QED
counterpart~\cite{sanchezmunoz14a}, but have instead a more flexible
temporal structure which could be further engineered
(or ``programmed'') through the decay rates of the
transitions. Similarly to the case of single photons, the cascaded
regime optimizes their correlations through temporal
liquefaction~\cite{zubizarretacasalengua24a}, showing that the
previous proposals~\cite{sanchezmunoz14a,bin20a} correspond to a
temporal gas of bundles. Interestingly, as an incoherent process, the
bundle purity (percentage of events featuring exactly the sought~$k$
number of photons in each bundle) is not fundamentally limited and
could thus be arbitrarily close to 1 regardless of the size of the
bundle.  This is therefore another example of the great flexibility
and versatility of the circular quantum cascade.  For bundling as for
Wigner crystallization, it provides much simpler, robust and tuneable
implementations.

\section{Two-Photon cascade in the Three-Level System}
\label{sec:Sat29Jun131254CEST2024}

The case~$N=3$ is both that of the simplest and most studied radiative
cascade under incoherent pumping. Largely, this is due to the rich
bi-exciton/exciton cascade in semiconductors~\cite{moreau01a,kiraz02a}
and its potential for generating entanglement~\cite{akopian06a}.  We
can expect the circular implementation to be also more easily realized
with fewer steps. The two steps with $N=3$ (if not counting the
excitation one) realize the paradigm of a photon cascade. For that
reason, we can tackle a more general study where we relax the
degeneracy between the parameters. This will allow us to get an idea
of how much has just been overlooked in such an approximation.  Before
that, however, it is instructive to pause and consider the case~$N=2$
which, as an incoherently pumped system, describe the archetypal
single-photon source (blue box in
Fig.~\ref{fig:Sat13Jul122521CEST2024}(a)).  Our formalism shows,
however, that it is the special case of the ``one-photon cascade in
the two-level system''. This adds to the well-known and standard
autocorrelation function
\begin{equation}
  \label{eq:Mon15Jul075735CEST2024}
  g_{1,1\{N=2\}}(\tau)=1-\exp\big(-(\gamma_0+\gamma_1)|\tau|\big)
\end{equation}
the cross-correlation between emission and excitation:
\begin{equation}
  \label{eq:Mon15Jul075823CEST2024}
  g_{1,0\{N=2\}}(\tau)=
  1+\left({\gamma_0\over\gamma_1}\right)^{\varsigma(\tau)}\exp\big(-(\gamma_0+\gamma_1)|\tau|\big)
\end{equation}
where~$\varsigma(\tau)\equiv {2\Theta(\tau)-1}$ is~$+1$ if~$\tau\ge 0$
and~$-1$ otherwise ($\Theta$ is the Heavyside function).  By circular
symmetry, one gets $g^{(2)}_{0,0}(\tau)=g^{(2)}_{1,1}(\tau)$ and
$g_{0,1}(\tau)=g_{1,0}(-\tau)$. This solution is notable for having
been apparently overlooked for so long. One of its merit is to clarify
why incoherent pumping of a two-level system fails to provide a good
single photon source: behind the
antibunching~(\ref{eq:Mon15Jul075735CEST2024}) lurks the strong
bunching~(\ref{eq:Mon15Jul075823CEST2024}) of the re-excitation, which
counteracts from its cascading nature, the suppression of coincidences
in the autocorrelations. This is the basic fact whose understanding
and mitigation should drive the design of efficient CW single-photon
sources. The $N=3$ steady-state solution of
Eq.~(\ref{eq:Thu25Jul181335CEST2024}) can now be found in full
generality:
\begin{equation}
  \label{eq:Fri29Dec183552CET2023}
  \vec p_\mathrm{ss}=\bigg(\sum_{i,j=0\atop i>j}^2\gamma_i\gamma_j\bigg)^{-1}
  \begin{pmatrix}
    \gamma_0\gamma_1\\
    \gamma_0\gamma_2\\
    \gamma_1\gamma_2
  \end{pmatrix}\,.
\end{equation}

The autocorrelation function is the same for both transitions, which is
also a feature of Eq.~(\ref{eq:Tue25Jun114035CEST2024}) as already
observed. Therefore, for~$i=1,2$:
\begin{multline}
  \label{eq:Sun30Jun104835CEST2024}
  g^{(2)}_{i,i}(\tau)=1+{}\\{}+{\gamma_{012}-\zeta\over2\zeta}e^{-{\gamma_{012}+\zeta\over2}|\tau|}
  -{\gamma_{012}+\zeta\over2\zeta}e^{-{\gamma_{012}-\zeta\over2}|\tau|}
\end{multline}
where we defined
\begin{equation}
  \label{eq:Sat29Jun142737CEST2024}
  \zeta\equiv\sqrt{\sum_{i=0}^2\gamma_i^2-\sum_{i,j=0\atop i\neq j}^2\gamma_i\gamma_j}
\end{equation}
as well as~$\gamma_{i_1\cdots i_k}=\gamma_{i_1}+\cdots+\gamma_{i_k}$
with as many indices as there are terms in the sum. We will use a bar
to denote a negative term,
$\gamma_{i_1\cdots
  \overline{\imath_k}}=\gamma_{i_1}+\cdots-\gamma_{i_k}$, so that,
e.g., $\gamma_{0\bar 12}\equiv\gamma_0-\gamma_1+\gamma_2$.
Here~$\gamma_0$ has the interpretation of a pumping rate.  For the
transition correlations in this case, we find:
\begin{widetext}
  \begin{multline}
    \label{eq:Sat29Jun132948CEST2024}
    g^{(2)}_{2,1}(\tau)=1+{}\\
    \begin{cases}{1\over4\zeta\gamma_2^2}\left[
      (\gamma_{0\bar 12}-\zeta)\big(\gamma_1^2+\gamma_2^2-(\gamma_0+\zeta)\gamma_{12}\big)e^{-{\zeta+\gamma_{012}\over2}\tau}-(\gamma_{0\bar 12}+\zeta)(\gamma_1^2+\gamma_2^2-(\gamma_0-\zeta)\gamma_{12})e^{{\zeta-\gamma_{012}\over2}\tau}\right] & \text{if~$\tau<0$\,,} \\
      {1\over 4\zeta\gamma_0\gamma_1}\left[(\gamma_{\bar 012}+\zeta)\big((\gamma_{0\bar
        1}+\zeta)\gamma_1+(2\gamma_0+\gamma_1)\gamma_2\big)e^{{\zeta+\gamma_{012}\over2}\tau}-(\gamma_{\bar
        012}-\zeta)\big((\gamma_{0\bar
        1}-\zeta)\gamma_1+(2\gamma_0+\gamma_1)\gamma_2\big)e^{-{\zeta-\gamma_{012}\over2}\tau}\right]
      & \text{if~$\tau>0$\,,}
    \end{cases}
  \end{multline}
\end{widetext}
while, following~Eq.~(\ref{eq:Tue9Jul141459CEST2024}),
$ g^{(2)}_{2,1}(\tau)=g^{(2)}_{1,2}(-\tau)$.

A first interesting observation from this general solution is to
consider the range of parameters that result in oscillations or, on
the other hand, in damped relaxations. This follows from the sign of
the radicand of Eq.~(\ref{eq:Sat29Jun142737CEST2024}), which
makes~$\zeta$ real or imaginary and its exponential in
Eqs.~(\ref{eq:Sun30Jun104835CEST2024})
and~(\ref{eq:Sat29Jun132948CEST2024}) an additional damping or an
oscillation, respectively. Spelling out the condition for
oscillations, we must
have~$2\gamma_0\gamma_1+2\gamma_0\gamma_2+2\gamma_1\gamma_2>\gamma_0^2+\gamma_1^2+\gamma_2^2$
with all~$\gamma_i>0$. Rearranging for~$\gamma_2$ as a function
of~$\gamma_0$ and~$\gamma_1$, this leads to
$\gamma_2^2-2\gamma_2(\gamma_0+\gamma_1)-(\gamma_0-\gamma_1)^2<0$.
The zeros of this quadratic equation thus give us the range
of~$\gamma_2$ that results in oscillations of the correlation
functions for a given~$\gamma_1$ and~$\gamma_2$ as
\begin{equation}
  \label{eq:Sun8Dec144133CET2024}
  (\sqrt{\gamma_1}-\sqrt{\gamma_2})^2<\gamma_0<(\sqrt{\gamma_1}+\sqrt{\gamma_2})^2\,.
\end{equation}
The same relations hold for any permutations of the indices. This
shows that if two rates are equal, oscillations occur as long as the
third (nonzero) rate is less than four times the common decay
rate. Clearly, the case where all three decay rates are equal satisfy
the condition for oscillations. While we do not provide an exact
statement for the general case, it seems clear that a small spread of
the decay rates favour oscillations while at the same time
accommodating large deviations for some of the parameters. With~$N=3$,
such oscillations are not resolvable from the correlation function in
a linear scale, since their amplitude is too small (they would be
visible in log-scale for~$g^{(2)}-1$), but this serves to illustrate
the different structural regimes that the mechanism supports and how
they survive or not different parameters. Extrapolated to larger~$N$,
liquid light should thus be accessible even with significant
inhomogeneities of the cascading steps.

The exact form~(\ref{eq:Sat29Jun132948CEST2024}) for the two-photon
cascade in a three-level system admits in regimes of interest simpler
expressions, that one can compare to the phenomenological two-photon
cascade correlations that assumes an uncorrelated (Poissonian) stream
of photons heralding another stream~\cite{zubizarretacasalengua23a}:
\begin{equation}
  \label{eq:Sun9Oct102241BST2022}
  g^{(2)}(\tau)=1+
  \begin{cases}
    (1-p)\displaystyle{\gamma_2\over\gamma_1}\exp(\gamma_2\tau) & \text{if~$\tau<0$},\\
    p\displaystyle{\gamma_2\over\gamma_1}\exp(-\gamma_2\tau) & \text{if~$\tau>0$}.\\
  \end{cases}
\end{equation}
where~$p$ describes the probability of good time ordering, $\gamma_1$
the emission rate of the uncorrelated heralding (first) photon and
$\gamma_2$ that of the heralded (second) photons. Namely, in the limit
of small excitation, i.e., when~$\gamma_0\ll \gamma_1,\gamma_2$, so
that~$\zeta\approx\sqrt{\gamma_1^2+\gamma_2^2-2(\gamma_1+\gamma_2)}=|\gamma_1-\gamma_2|$,
in which case Eq.~(\ref{eq:Sat29Jun132948CEST2024}) takes the form:
\begin{equation}
  \label{eq:Sat29Jun144049CEST2024}
  g^{(2)}_{2,1}(\tau)=1+
  \begin{cases}
    -\exp(\gamma_1\tau) & \text{if~$\tau<0$}\,,\\
    {\gamma_2\over\gamma_0}\exp(-\gamma_2\tau) & \text{if~$\tau>0$}\,,
  \end{cases}
\end{equation}

\begin{figure}
  \includegraphics[width=.66\linewidth]{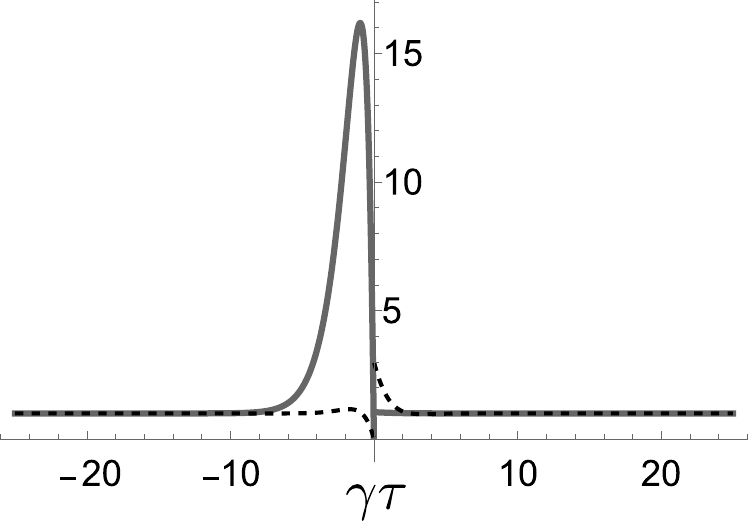}
  \caption{Two-photon cascade in a three-level system with unbalanced
    rates (thick solid line,
    with~$(\gamma_1,\gamma_2)/\gamma_0=(1.1,0.025)$), resulting in a
    markedly different correlation function (compare with all rates
    equal to the mean~$\gamma\approx 0.71\gamma_0$ as the dashed thin
    line). The heralded photon is uncorrelated with the heralding one,
    while the opposite transition is strongly bunched at small
    negative delays, though still obeying the
    constrain~(\ref{eq:Sat29Jun150141CEST2024}) at~$\tau=0$. Similar
    strong departures are expected for the general~$N$ case. These
    parameters satisfy Eq.~(\ref{eq:Sun8Dec144133CET2024}) so the
    functions are forever oscillating, although this is not visible
    on a linear
    scale. \label{fig:Wed31Jul012741CEST2024}}
\end{figure}

On the other hand, in the limit of
high-pumping~$\gamma_0\gg\gamma_1,\gamma_2$, one finds:
\begin{multline}
  \label{eq:Sat29Jun145429CEST2024}
  g^{(2)}_{2,1}(\tau)=1+\\
  \begin{cases}
    {\gamma_1\over\gamma_2}\exp([\gamma_1+\gamma_2]\tau)-\left(1+{\gamma_1\over\gamma_2}\right)\exp(\gamma_0\tau) & \text{if~$\tau<0$}\,,\\
    {\gamma_2\over\gamma_1}\exp(-[\gamma_1+\gamma_2]\tau) & \text{if~$\tau>0$}\,.,
  \end{cases}
\end{multline}
The second (negative) term for~$\tau<0$ is negligible for most~$\tau$
except very close to zero where it forces the correlations to be
exactly antibunched. This is the main deviation of the cascading
scheme as compared to Eq.~(\ref{eq:Sun9Oct102241BST2022}) which
assumes uncorrelated heralders. The cascade, on the other hand,
requires the same excitation to go up and down the ladder and thus
demands that
\begin{equation}
  \label{eq:Sat29Jun150141CEST2024}
  \lim_{\tau\to0\atop\tau<0}g^{(2)}_{2,1}(\tau)=0
\end{equation}
which is the counterpart of Eq.~(\ref{eq:Sat13Jul153352CEST2024})
satisfied also by Eq.~(\ref{eq:Sat29Jun132948CEST2024}) and the
low-driving approximation~(\ref{eq:Sun9Oct102241BST2022}).  Since, on
the other hand
\begin{equation}
  \label{eq:Sat29Jun152012CEST2024}
  \lim_{\tau\to0\atop\tau>0}g^{(2)}_{2,1}(\tau)=1+\left({1\over\gamma_0}+{1\over\gamma_1}\right)\gamma_2
\end{equation}
(with limits~$1+{\gamma_2\over\gamma_0}$ for
low~(\ref{eq:Sat29Jun144049CEST2024}) and~$1+{\gamma_2\over\gamma_1}$
for high~(\ref{eq:Sat29Jun145429CEST2024}) pumping, respectively),
there is again the discontinuity at~$\tau=0$, as is also the case in
the phenomenological model and in the more general
cascade~(\ref{eq:Tue9Jul155646CEST2024}).  This would be resolved with
a photo-detection theory~\cite{delvalle12a}. Lifting the degeneracy of
the relaxation rates show that the
discontinuity~(\ref{eq:Tue9Jul155646CEST2024}) is the smallest that
can be, and that large~$\gamma_2$ or small~$\gamma_0$
and/or~$\gamma_1$ result in strong discontinuities, as the cascade is
rarefied as compared to its Poisson occurence. It is interesting, in
this regard, to consider the opposite limit which tames down the
correlations, as shown in Fig.~\ref{fig:Wed31Jul012741CEST2024}, where
one sees that the decorrelation of the heralded photon in the cascade,
comes at the cost of strong correlations in the wrong order, i.e., of
the heralding one instead, despite the perfect suppression of
coincidences from Eq.~(\ref{eq:Sat29Jun150141CEST2024}). This realizes
a ``backward-cascade'' where the detection of photons from the
supposedly ``heralding'' transition tells us nothing about the
subsequent transition of an heralded one, which are thus detected
randomly in time, while such a transition effectively ``heralds its
heralder'', i.e., the photon from the prior transition is in fact
emitted shortly and strictly afterwards (in the wrong time order and
never at the same time). That is because it is, of course, the next
heralder, but this fact does not transpire in the emission and such a
``feature'' would not be easy to achieve otherwise. Reversing the
detectors with Eq.~(\ref{eq:Tue9Jul141459CEST2024}), one thus has
uncorrelated heralders heralding slightly delayed photons. We thus
have a recipe to implement closely the phenomenological cascade of
Eq.~(\ref{eq:Sun9Oct102241BST2022}), would that be requested.

Finally, would we consider the~$N=3$ case not as a two-photon cascade
from the three-level system, but as the generic circular cascade of
Fig.~\ref{fig:Sat13Jul122521CEST2024}(b) where all transitions can be
correlated, then $g^{(2)}_{0,0}$ is given by
Eq.~(\ref{eq:Sun30Jun104835CEST2024}) and the cross-correlations are
similarly given by Eq.~(\ref{eq:Sat29Jun132948CEST2024}) with rotation
of the parameters, namely, $g^{(2)}_{1,0}$ is obtained from the
substitutions~$(\gamma_0,\gamma_1,\gamma_2)\to(\gamma_2,\gamma_0,\gamma_1)$
and~$g^{(2)}_{0,2}$ from the
substitutions~$(\gamma_0,\gamma_1,\gamma_2)\to(\gamma_1,\gamma_2,\gamma_0)$,
while the other orders follow from
Eq.~(\ref{eq:Tue9Jul141459CEST2024}).  These correlations behave
qualitatively as discussed in Fig.~\ref{fig:Sat13Jul122521CEST2024}
but lifting the degeneracy of the transition rates and with similar
enhancement and distortions of their correlations. This shows that
non-degenerate decay rates do not enlarge the number of possible
traces.

\section{Discussion and Conclusions}

We have provided a theory of circular quantum cascades, whereby a
level can be excited from one direction only. While this seems to
differ little from widespread and thoroughly studied two-way cascades,
where the state can be excited from ``above'' (by relaxation) or from
``below'' (by excitation), this actually presents a considerably
richer dynamics, able to endow the system with strong correlations
that are washed away in other cases by breaking down the ladder into a
succession of pairwise cascades. Indeed, although the formalism to
describe circular cascades brings little novelty from those developped
to describe other types, the nature of the solutions---developing
all-time oscillations under incoherent excitation of the system---make
a dramatic departure from the mere relaxation regimes previously
discussed. For the simplest case of~$N=3$ for which we provided an
exact description with no loss of generality, it might be possible to
revisit existing platforms that realize a close-enough scenario of
one-way cascading, and, in the previously unsuspected regime of
Eq.~(\ref{eq:Sun8Dec144133CET2024}), chase the onset of photon
liquefaction. In the general case where~$N$ can be made arbitrarily
greater, we provided a  comprehensive description of all
the possible traces for the various combinations of transitions that
can be correlated (which to the best of our knowledge remains to be
similarly classified for traditional cascades). There, we
highlighted how autocorrelations surprisingly recover highly-sought
regimes of photon phase transitions, such as those realized in large
ensemble of coupled optical
cavities~\cite{hartmann06a,angelakis07a,greentree06a}. These
implementations have been described as a ``complex architecture [that]
represents a considerable experimental challenge'' by an EIT Rydberg
group seeking similar correlations~\cite{chang08a}, also praising the
Rydberg out-of-equilibrium character and its facility to transfer them
to the optical field. Our mechanism can make the same remarks on the
EIT Rydberg platform: the cascaded system is much more straightforward
experimentally, being in principle available with a single multi-level
emitter, it also needs no equilibrium, in fact not even coherent
driving, and cares little about underlying details of the excitation,
and it is directly and intrinsically built into the optical
field. This challenges the previously held view that the ``formation
of a Tonks--Girardeau gas of photons is fundamentally a collective
many-body effect''~\cite{chang08a}. This in fact emerges as a much
more fundamental and universal feature of the optical field itself,
regardless of the underlying mechanism which produces it, that we
identify as that of good single-photon emission.  Nevertheless, the
coupled cavities and Rydberg EIT physics being deeply rooted indeed in
strongly-correlated many-body physics---whose photon correlations are
directly linked to the pair correlation function of the Lieb-Liniger
gas~\cite{muth10a}, where they are interpreted as Friedel
oscillations~\cite{friedel58a}---this gives further credence to our
earlier suggestions of a condensed-matter, thermodynamically inspired
description of the optical field~\cite{zubizarretacasalengua24a}. In
our case, photons are not interacting, they merely inherit or imprint
correlations that are those of a liquid, so even the denomination of
``liquid light'' might not be entirely correct and one should instead
speak of \emph{liquid time}, since this is properly, and maybe
exclusively, the ticks in time that exhibit the features, not its
carriers. At this stage, the connection with Wilczek's variation of
his time crystal, which he dubbed \emph{time
  liquid}~\cite{wilczek19a}, becomes irresistible, and we believe
indeed that, despite lacking here too the many-body and
strongly-interacting characters typically attached to these ``time
materials'', our results provide an interesting platform from which
local time-ordering spontaneously emerges, as a result of breaking the
bidirectionality of excitation transfer. Interestingly, this also
brings into that discussion the Glauber two-photon correlation
function which, despite its ubiquity in quantum optics, is not, to the
best of our knowledge, directly considered as a quantifier of quantum
optical coherence in the description of time crystals (we find no
mention of bunching or antibunching neither in the leading
publications nor the several
reviews~\cite{sacha17a,khemani_arXiv19a,hannaford22a,zaletel23a} or
comments~\cite{coleman13a,gibney17a,wilczek19a}, and even a
textbook~\cite{sacha_book20a} on this topic). Such conceptual bridges
between hitherto disconnected disciplines should benefit all of them:
in quantum optics, it precises the meaning of a perfect single-photon
source and suggests mechanisms to implement them; in condensed-matter,
it enlarges the realm of
platforms 
with a purely photonic one, possibly with no spatial extension, as
well as providing the liquid phase of those emerging time materials.

Another striking demonstration of the might of circular
cascades is their built-in ability to generate CW $N$-photon bundles,
for any integer~$N$ and with no fundamental restriction on the bundle
purity, which appear to also provide considerable improvement on their
cavity QED counterparts. At any rate, proper attention to such
potentialities and thus the exploitation of the wonders they promise,
will require the experimental feasibility of such cascades. It is
beyond the scope of this text to devise a microscopic system to
achieve that, although we reiterate that in our view, despite the
physical constraints on relaxation rates in~$N$-level quantum
systems~\cite{schirmer04a}, this seems a much simpler task than those
involving strongly-correlated many-body platforms driven in extreme
regimes. We can imagine several ``poor man'''s solutions to the
problem, including pulsed excitation to avoid re-excitation or
coherent driving to isolate the~$\gamma_0$ transition from the
others. In such cases, however, there would be the risk of mistaking
the oscillations for those imparted by the driving
itself~\cite{cheng06b,agarwal77a}. With the onset of chiral quantum
optics~\cite{lodahl17a} and topological photonics~\cite{ozawa19a}, we
have no doubt that resourceful inventors can find a faithful
implementation of a genuine one-way one-photon (or one-quantum, in
other platforms) cascade, with the result of producing not one, but a
variety of stationary quantum light, yet so strongly correlated that
they locally appear to be pulsed.

From our side, rather than indulging into such designs, we wish to
conclude with the reiterated observation that the current theory of
circular cascades, as well as, incidentally, those describing other
types of cascades, rely on correlating transitions, as opposed to
correlating detected photons. The seminal cascade from resonance
fluorescence~\cite{aspect80a} has already been upgraded along those
lines~\cite{zubizarretacasalengua23a} and unraveled a plethora of
unsuspected features. We believe that much physics and further
engineering await to be revealed by similarly applying the sensor
formalism~\cite{delvalle12a} to the general problem of photon
cascades---circular or traditional---with the effect of removing
undesirable artifacts such as discontinuities in the correlation
functions, taking more seriously photon indistinguishability as well
as interferences, and, not least, providing a complete leapfrog
(off-peak) picture of the transitions~\cite{sanchezmunoz15a}. By
placing the observation at the heart of the
process~\cite{zubizarretacasalengua24b}, one could show that quantum
jumps and quantum cascades are two faces of the same coin.  Thanks to
their generality and rich features, both conceptually (time liquids)
and phenomenologically (programmable multiphoton sources), circular
quantum cascades should provide a particularly fruitful theoretical
playground in such endeavors.

\begin{acknowledgments}
  We thank Drs.~Arturo Camacho Guardian, Asaf Paris Mandoki, Giuseppe
  Pirucci and Hugo Alberto Lara Garcia for attracting our attention to
  the Rydberg blockade correlations, at the occasion of the first Etic
  Tlahuilli school. EdV and FPL thank Guillermo del Valle and Lotte
  Crombaghs for their hospitality on their wedding week, during which
  this work was finalized.  EdV acknowledges support from the TUM-IAS
  Hans Fischer Fellowship, the CAM Pricit Plan (Ayudas de Excelencia
  del Profesorado Universitario), Sinérgico CAM project Y2020/TCS-6545
  (NanoQuCo-CM) and the Spanish Ministry of Science, Innovation and
  Universities under grant AEI/10.13039/501100011033 (2DEnLight) and
  through the ``María de Maeztu'' Programme for Units of Excellence in
  R\&D (CEX2023-001316-M).  FPL acknowledges support from the
  HORIZON~EIC-2022-PATHFINDERCHALLENGES-01 HEISINGBERG project
  101114978.
\end{acknowledgments}

\bibliographystyle{naturemag}
\bibliography{sci,Books,arXiv}

\end{document}